\documentstyle[10pt,aaspp4,psfig]{article}

\def\lap{\lower.5ex\hbox{$\; \buildrel < \over \sim \;$}}
\def\gap{\lower.5ex\hbox{$\; \buildrel > \over \sim \;$}}
\def\dcl{d_{\rm cl}}
\def\lb{\langle}
\def\rb{\rangle}
\def\dt{{\textstyle{\left({\delta T\over T}\right)}}}
\def\dtn{{\textstyle{{\delta T\over T}}}}
\begin{document}

%\baselineskip = 0.5\baselineskip

\title 
{Small-scale anisotropy of the cosmic background radiation \\
and scattering by cloudy plasma}
\author{P. J. E. Peebles\altaffilmark{1}
and R. Juszkiewicz\altaffilmark{2}}

\altaffiltext{1}{Department of Physics, 
Princeton University, 
Princeton, NJ 08544; \\
pjep@pupgg.princeton.edu}
\altaffiltext{2}
{Copernicus
Center, Bartycka 18, 00 716 Warsaw, Poland;
roman@camk.edu.pl}

\begin{abstract}

If the first stars formed soon after decoupling of baryons 
from the thermal cosmic background radiation (the CBR) the radiation 
may have been last scattered in a cloudy plasma. 
We discuss the resulting small-scale anisotropy of the CBR 
in the limit where the plasma clouds are small
compared to the mean distance between clouds along a line of sight. 
This complements the perturbative analysis
valid for mildly nonlinear departures from homogeneity 
at last scattering. We conclude
that reasonable choices for the cloud parameters imply CBR anisotropy 
consistent with the present experimental limits, in agreement 
with the perturbative approach. This means the remarkable 
isotropy of the CBR need not contradict the early small-scale 
structure formation predicted in some cosmogonies. 

\end{abstract}


\section{Introduction}
\label{sec-intro}

Cosmogonies in which the baryons were concentrated in clouds 
at the epoch of last scattering of the thermal cosmic background 
radiation (the CBR), as in isocurvature
models (\cite{pe94}, \cite{pe97a}), are in qualitative agreement with
the appearance in quasar absorption line spectra of a
well-advanced state of structure formation at redshift $z\sim 5$, 
but the large amplitude of the density fluctuations may produce
significant small-scale anisotropy in the CBR. The analysis of the CBR
anisotropy in second order perturbation theory was introduced by 
\cite{os86}. Vishniac's (1987) more detailed 
investigation has been confirmed by \cite{hu94}, 
\cite{do95}, and \cite{hu97}.
The analysis and numerical
evaluation are extended in \cite{pe95} and Persi et al. 
(1995). The application of perturbation theory 
may be uncertain if early structure formation produced highly nonlinear 
departures from homogeneity at the epoch of last scattering, however.
To investigate this we have developed a nonperturbative model 
for scattering in a cloudy distribution of plasma. The expression for  
the CBR anisotropy $\delta T/T$ in the strongly cloudy limit bears a close
resemblance to the Ostriker-Vishniac relation, and the numerical
results for $\delta T/T$ accordingly are similar.  

Our analysis, which extends previous discussions by  
\cite{ho89}, \cite{pe90}, \cite{ag96} and \cite{gr98}, assumes the CBR was 
last scattered by free electrons in clouds with density contrast
well above unity, so the mean free distance $t_f$ between
intersections of gas clouds along a line of sight is much larger
than the typical cloud size $\dcl$. The simplification offered by this
limit is that the details of the matter distribution and motion on
scales from $\dcl$ to $t_f$ are not important, because a line of
sight on average samples only one cloud over the distance $t_f$. 
Thus we can model the clouds as an inhomogeneous random Poisson
process. In the process the joint distribution of cloud motions 
and scattering optical depths  
as a function of position along the line of sight is determined
by the plasma density and velocity fields averaged through a window of
width $t_f$. If $t_f$ is larger than the scale of non-linear density 
fluctuations then the CBR anisotropy $\delta T/T$ is the sum in 
quadrature of a perturbative contribution and a shot noise term.

The next section shows the relation between the shot noise term in 
$\delta T/T$ and the Ostriker-Vishniac (1986) effect. The model of
the clouds as an inhomogeneous random Poisson process is presented 
in \S 3, and a simplified treatment of the effect of correlated
cloud motions is discussed in \S 4. In \S 5 we present numerical
examples of the expected CBR anisotropy.  

\section{Shot noise and the Ostriker-Vishniac anisotropy}
\label{noise-term}

\cite{os86} showed that the CBR anisotropy at angular resolution $\theta$ 
produced by the peculiar motion of the plasma at last scattering at
proper expansion time $t_s$ is
\begin{equation}
	\dt _\theta ^2 \; \; \sim \; \;
	v_\theta {}^2 \delta _\theta {}^2{r_\theta\over t_s}\, .
\label{eqn-ov1}
\end{equation}
Here $r_\theta$ is the proper length subtended at the epoch of last
scattering by the observed angle $\theta$, and
$\delta _\theta$ and $v_\theta$ are the rms values of the density
contrast and the line of sight component of the
peculiar velocity averaged through a window of width $r_\theta$. 
Ostriker \&\ Vishniac based this relation on the lowest nontrivial
order of perturbation theory. When the CBR is scattered in discrete
and well-isolated clouds of plasma the analogous expression is
obtained as follows.  

Suppose the CBR is scattered in clouds with typical optical depth
\begin{equation}
	\tau \; = \; \sigma\Sigma _e\, ,
\label{eqn-ov2}
\end{equation}
where $\sigma$ is the Thomson cross section and $\Sigma _e$ is
the characteristic free electron column density of a cloud. 
If $\tau\lap 1$ the CBR temperature perturbation observed at angular 
resolution smaller than a cloud size may be approximated as 
\begin{equation}
	\dtn \; \sim \; \int _{t_s}\sigma n_e v dt
	\; \sim \; \sum\tau _\alpha v_\alpha\, .
\label{eqn-ov3}
\end{equation}
In the integral $n_e$ and $v$ are the free electron number density and
streaming velocity as functions of position along a line of sight, and
the integral is over a proper distance comparable to the expansion time 
$t_s$ at the epoch where the universe last is optically thick,
$\sigma n_et_s\sim 1$. (Here and below the velocity of light set
to unity.) The sum is over $N\sim\tau ^{-1}\sim t_s/t_f$ clouds,
where $t_f$ is the mean distance between cloud intersections
along a line of sight. 
If $t_f$ is larger than the nonlinear clustering length then the sum 
in equation adds as a random walk, and the mean 
square temperature perturbation along a line of sight is
\begin{equation}
	\dt^2 \;\; \sim \;\; v^2\tau ^2N \; \sim \; v^2\tau\, .
\label{eqn-ov4}
\end{equation}
If the CBR is observed in a beam with angular size $\theta$ larger 
than the angular diameter $\theta _{\rm cl}$ characteristic of clouds 
with proper width $d_{\rm cl}$ the mean square anisotropy is 
\begin{equation}
	\dt_\theta ^2 \;\; \sim \;\; 
	v^2\tau {\theta _{\rm cl}^2\over\theta ^2} \;
	\sim \;\; v^2\tau\left(\dcl\over r_\theta\right) ^2\, ,
\label{eqn-ov5}
\end{equation}
because we are ignoring correlations among cloud positions and
motions. As in equation~(\ref{eqn-ov1}), $r_\theta$ is the proper length 
subtended by the beam size $\theta$ at last scattering. 
The number of free electrons in a cloud  is $\sim\Sigma _e\dcl ^2$, 
so the number density $n_{\rm cl}$ of clouds satisfies
\begin{equation}
	\Sigma _ed_{\rm cl}^2n_{\rm cl} \;\; \sim \;\; \bar n_e
	\; \sim \; 1/(t_s\sigma )\, , 
\label{eqn-ov6}
\end{equation}
where $\bar n_e$ is the large-scale mean number density of free
electrons. Equations~(\ref{eqn-ov5}) and (\ref{eqn-ov6}) give
\begin{equation}
	\dt_\theta ^2 \; \sim \; 
	v^2\delta _\theta ^2{r_\theta\over t_s}\, ,
	\qquad\qquad    
        \delta _\theta ^2 \; = \; {1\over n_{\rm cl}r_\theta ^3}\, .
\label{eqn-ov7}
\end{equation}
This is the same form as the Ostriker-Vishniac 
expression in equation (\ref{eqn-ov1}), where $v$ is the rms 
line-of-sight peculiar
streaming velocity of the material within a cloud and 
$\delta _\theta$ is the shot noise contribution to the rms value 
of the fractional fluctuation in the number of clouds found within 
a randomly placed sphere of width $r_\theta$.

In the next section we generalize this analysis to an
inhomogeneous random Poisson process that takes account of the
large-scale correlations in cloud positions and motions and the 
cosmic evolution of the cloud parameters. The former have
little effect on the small-scale CBR anisotropy, for \cite{su78}
noted that the CBR averages the perturbations of the
peculiar motions across the Hubble length $t_s$ at the 
epoch of last scattering, suppressing the contribution to the 
anisotropy. He also showed that the contribution is 
further suppressed by the anticorrelation of the peculiar motions 
on either side of a density fluctuation. This effect was
later rediscovered and confirmed by \cite{ka84}.
 
\section{Poisson model for the scattering clouds}
\label{poisson-process}
\subsection{The inhomogeneous Poisson process}
\label{radiative-transfer}
A suitable approximation to the transfer equation for the CBR
temperature perturbation observed at epoch $t_o$ along a line of
sight through a given electron distribution $n_e$
and streaming velocity field with a line-of-sight component
$v$ is
\begin{equation}
	\dtn \; = \; \int _o^{t_o} \sigma n_ev\, dt
	\, \exp -\int _t^{t_o}\sigma n_edt'\, .
\label{eqn-pp1}
\end{equation}
This expression ignores the gravitational perturbation to the
CBR, which is small in the small angle limit of interest here,
and the inhomogeneity in the space distribution of the radiation
at last scattering, for the observations tell us  
$\delta T/T\sim 10^{-5}$ at last scattering, while the velocity
term in equation (\ref{eqn-pp1}) is $v/c\sim 10^{-3}$. It also ignores the 
relativistic correction of order $\sim (v/c)^2$ discussed by 
\cite{hu94} and \cite{do95}, for in the
small and relatively cool clouds to be expected at high redshift 
this correction is on the order of $v/c\sim 10^{-3}$ times the 
Ostriker-Vishniac term. (The $(v/c)^2$ term is induced by the 
velocity-velocity nonlinear coupling, while the perturbative
analog of the above equation comes from the
velocity-density coupling).

We are assuming the CBR is last scattered in well-separated
clouds of plasma. If the optical depth of a cloud on the line
of sight is $\tau$, the probability the cloud has scattered a
particular photon into the beam is $1-e^{-\tau}$. If the peculiar
velocity of the cloud along the line of sight is $v$ the
contribution to the integral over $t$ in equation~(\ref{eqn-pp1}) by
this cloud is  
\begin{equation}
	\dt _{\rm cl} \; = \; (1-e^{-\tau })\, v
	\,\exp -\int _t^{t_o}\sigma n_edt'\, ,
\label{eqn-pp2}
\end{equation}
where the integral runs over all the plasma intersected by the
line of sight subsequent to this cloud. Thus in our
cloud model we rewrite equation~(\ref{eqn-pp1}) as
\begin{equation}
	\dtn \; = \; \sum_{iab}n_{iab}(1 - e^{-\tau _a})v_b
	\,\exp -\sum_{k>i,cd}n_{kcd}\tau _c\, ,
\label{eqn-pp3}
\end{equation}
where $n_{iab}=1$ if there is a cloud with optical depth in the
range $\tau _a$ to $\tau _a+d\tau _a$ and peculiar velocity in the
range $v_b$ to $v_b+dv_b$ at position $t_i$ to $t_i+dt_i$ along the
line of sight, and $n_{iab}=0$ otherwise. 

Equation~(\ref{eqn-pp3}) represents the strongly nonlinear
cloud-like distribution of the scattering plasma 
by the inhomogeneous random
Poisson  process $n_{iab}$. The inhomogeneity of the process
includes the time evolution of the mean cloud properties and the
large-scale structure in the distribution and motion of the
clouds. Thus we compute statistical averages 
in two steps: first average over the Poisson process for
given smoothed fields that represent the large-scale structure, 
and then use perturbation theory to compute the average over
the ensemble of smoothed fields. 

For a given realization of the 
smoothed fields the expectation value of the Poisson
process $n_{iab}$ in equation~(\ref{eqn-pp3}) is   
\begin{equation}
	\lb n_{iab} \rb = f_{iab}d\tau_a dv_b dt_i /t_f(i) \, .
\label{eqn-pp4}
\end{equation}
The probability of finding a cloud at the position labeled by
$t_i$ in the interval 
$dt_i$ is $dt_i/t_f$, where $t_f(i)$ is the local mean free
distance between cloud intersections in the given large-scale
structure. If there is a cloud at $t_i$ the probability distribution 
in its optical depth and velocity
is $f_{iab}$, with the normalization $\int f_{iab}d\tau _adv_b=1$. 
Thus the average over the Poisson process of the optical depth
of a cloud found at point $t_i$ on the line of sight is
\begin{equation}
	\langle\tau\rangle _i = \int f_{iab}\tau _ad\tau _adv_b =
	\sigma n_e(i)t_f(i)\, .
\label{eqn-pp5}
\end{equation}
In the last expression the mean plasma density $n_e(i)$ 
is a function of position $t_i$ along the line of sight, and 
averaged over the mean free path $t_f(i)$. If the mean optical 
depth $\lb\tau\rb _i$ is independent of position this equation 
just says the smoothed electron density varies inversely as 
the mean free distance between clouds. 

The mean value of the product of the electron
density and streaming velocity $v(i)$ at $t_i$ is
\begin{eqnarray}
\langle n_e v \rangle  & =  & {1 \over \sigma dt_i}\sum_{ab}
		              \langle n_{iab}\rangle \tau_a v_b \nonumber \\
	               & =  & {1\over\sigma t_f(i)} \;
                               \int d\tau_a dv_b\tau_a v_b
		              f_{iab} \nonumber \\
	               & =  & n_e(i) \; \; {\langle\tau v\rangle_i
			\over\langle\tau\rangle_i} \, .
\label{eqn-pp6}
\end{eqnarray}
The last line, which uses equation~(\ref{eqn-pp5}), expresses the average of
$n_ev$ across the Poisson process as the product of the smoothed electron
density field and the smoothed cloud velocity field weighted by the cloud
optical depth. 

\subsection{The variance of the CBR temperature}
\label{dT/Tvariance}

In the computation of the mean of $(\delta T/T)^2$ we consider first 
the average over the Poisson process for a given
realization of the smoothed fields that describe the large-scale
structure of the plasma distribution. 

Because $n_{iab}$ in
equation~(\ref{eqn-pp4}) is either zero or unity the mean of any positive power 
of $n_{iab}$ is the same as the mean of $n_{iab}$. By expanding
the exponential as a power series, one sees that 
\begin{equation}
	\lb e^{-n_{iab}\tau_a}\rb = 
	1 - \lb n_{iab}\rb (1 - e^{-\tau_a})\, .
\label{eqn-pp7}
\end{equation}
Averages across the Poisson process may be computed independently
for each different cell label $iab$, so the expectation
value of the visibility function in equation~(\ref{eqn-pp3}) is 
\begin{eqnarray}
\lb e^{-\sum n_{iab}\tau_a}\rb
	& = & \prod _{iab}\lb e^{-n_{iab}\tau_a}\rb \nonumber \\
	& = & \prod\left[1 - \lb n_{iab}\rb 
		\left(1 - e^{-\tau_a}\right)\right]\nonumber \\
	& = & \exp -\int dt_i(1 - \lb e^{-\tau}\rb_i)/t_f(i)
			 \, .
\label{eqn-pp8}
\end{eqnarray}
In the last line, the term $e^{-\tau _a}$ is averaged over the
distribution $f_{iab}$ of optical depths for clouds
found at position $i$ along the line of sight, as in
equation~(\ref{eqn-pp5}).  If the cloud optical depths are large, so 
$\lb e^{-\tau }\rb\ll 1$, equation~(\ref{eqn-pp8}) is just the
probability the line of sight intersects no clouds in the Poisson
process. If  $\lb\tau\rb\ll 1$ equation~(\ref{eqn-pp8}) is the usual
expression in perturbation theory, 
\begin{equation}
	\lb e^{-\sum n_{iab}\tau_a}\rb = 
	e^{-\int dt\lb\tau\rb /t_f} =
	e^{-\int \sigma n_edt}\, .
\label{eqn-pp9}
\end{equation}
Here $n_e$ is the local smoothed free electron number density defined
in equation~(\ref{eqn-pp5}). 

The average of the transfer equation~(\ref{eqn-pp3}) across the Poisson
distribution for a given realization of the large-scale structure 
is
\begin{equation}
	\lb  \, \dtn \, \rb \; = \; \int_0^{t_o} {dt\over t_f(t)}
	\; \lb (1-e^{-\tau })v\rb_t \, \exp -\int_t^{t_o}
	{dt'\over t_f(t')} \left( 1-\lb e^{-\tau}\rb_{t'}\right)\, .
\label{eqn-pp10}
\end{equation}
Here again the averages are over the distribution $f_{iab}$ of
optical depths and line of sight velocities of clouds found at
a given position $t$ along the line of sight. 
If the cloud optical depths are small, we can
use the expression for the smoothed electron number density 
$n_e(t)$ in equation~(\ref{eqn-pp5}) to rewrite 
equation~(\ref{eqn-pp10}) as 
\begin{equation}
	\lb \, \dtn \, \rb \; = \; \int _0^{t_o} 
	\sigma n_e dt\, v(t)\, e^{-\int_t^{t_o}\sigma n_edt'}\, ,
  \qquad v(t) \; = \; \lb\tau v\rb_t/\lb\tau\rb_t\, .
\label{eqn-pp11}
\end{equation}
This is the form one would write down in perturbation theory,
where the smoothed field $v(t)$ as a function of position $t$
along the line of sight is the mean peculiar velocity weighted by
the cloud optical depth, as in equation~(\ref{eqn-pp6}).

In the expression for the mean square value of the temperature
perturbation along a line of sight we have to consider separately
the squared terms and the cross terms from  
the sum over the space position index $i$ in 
equation~(\ref{eqn-pp3}).
Thus we write the average over the Poisson process as
\begin{equation}
	\lb \, \dt^2 \, \rb \; = \; \dt^2_s
		+ \dt^2_c\, .
\label{eqn-pp12}
\end{equation}
As we now discuss, the squared terms generalize the shot noise 
in equation~(\ref{eqn-ov4}) and the cross
terms approximate the variance of $\delta T/T$ in
perturbation theory for the smoothed fields. 

The sum of the squared terms in equation~(\ref{eqn-pp3}) is
\begin{equation}
	\dt ^2_s = \sum_{iabcd}
	\lb n_{iab}n_{icd}(1-e^{-\tau_a})
	(1-e^{-\tau _c}) v_bv_d
	\prod_{k>i,ef}e^{-2n_{kef}\tau_e}\rb\, .
\label{eqn-pp13}
\end{equation}
In the last product we have taken account of the fact that each
exponential factor $e^{-n_{kef}\tau _e}$ appears twice, once from
each factor of $\delta T/T$. As in equation~(\ref{eqn-pp8}) we can compute
separately the average over the Poisson process for each different 
cell $iab$. Also, we have
\begin{equation}
	\lb n_{iab}n_{icd}\rb = \lb n_{iab}\rb 
	\delta_{ac}\delta_{bd}\, ,
\label{eqn-pp14}
\end{equation}
because the terms where $a\not= c$ or $b\not= d$ are of second
order in $dt_i$. Thus equation~(\ref{eqn-pp13}) is
\begin{equation}
	\dt ^2_s  \; = \;
	\int_0^{t_o}{dt\over t_f}\lb v^2(1-e^{-\tau })^2\rb_t
	\exp -\int _t^{t_o}{dt'\over t_f(t')}
	\left( 1 - \lb e^{-2\tau }\rb _{t'}\right)\, .
\label{eqn-pp15}
\end{equation}
We complete the discussion of this shot noise term in \S 3.3, after 
dealing with the cross terms in equation~(\ref{eqn-pp12}). 

The sum of the cross terms in equation~(\ref{eqn-pp3}) is
\begin{eqnarray}
	\dt ^2_c & = & 2\sum _{i<j}\sum_{abcd}
	(1-e^{-\tau_a})(1-e^{-\tau _c}) 
        v_b v_d \lb n_{iab}\rb \nonumber \\
	& \times &\lb n_{jcd} \prod_{ef} e^{-n_{jef}\tau_e}\rb
	\prod_{i<k<j}\prod_{gh}\lb e^{-n_{kgh}\tau_g}\rb
	\prod_{l>j,mn}\lb e^{-2n_{lmn}\tau_m}\rb\, .
\label{eqn-pp16}
\end{eqnarray}
The first expectation value in this equation refers to the earlier 
time $t_i$ along the line of sight. The second expectation value 
contains the exponential factors at the later time $t_j$ along the
line of sight. In the next group of expectation values the exponential
factors are evaluated at times between $t_i$ and $t_j$. In the last
group, at times greater than $t_j$, each exponential factor appears 
twice, from the two factors of $\delta T/T$. We have
\begin{equation}
	\lb n_{jcd}\prod _{ef}e^{-n_{jef}\tau_e}\rb =
	\lb n_{jcd}e^{-n_{jcd}\tau_c}\rb = 
	\lb n_{jcd}\rb e^{-\tau_c}\, .
\label{eqn-pp17}
\end{equation}
The first step follows because the exponential factors with
$ef$ not equal to $cd$ introduce terms of second order in
$dt_j$. The last step follows by the argument used in 
equation~(\ref{eqn-pp7}). With equations~(\ref{eqn-pp4}) and 
(\ref{eqn-pp8}) we get
\begin{eqnarray}
	\dt ^2_c & = & 2\int_{i<j}
	\; \; {dt_idt_j\over t_f(i)t_f(j)}\, I \, J\, , \nonumber \\
	I & = & \lb\lb(1-e^{-\tau })v\rb_i \lb
	(1-e^{-\tau })e^{-\tau}v\rb_j\rb\, ,\nonumber\\
	J & = & e^{-\int_i^jdt_k(1-\lb 
	e^{-\tau }\rb) /t_f}
	e^{-\int _{l>j}dt_l(1-\lb e^{-2\tau }\rb) /t_f}\, .
\label{eqn-pp18}
\end{eqnarray}
In this equation the outermost brackets, $\lb\ldots\rb$,
denote an average over an ensemble of large-scale velocity fields,
while the inner brackets, $\lb\ldots\rb_i$, denote an
average over the Poisson process for each cell $iab$ for a given
realization of the ensemble of velocity fields.

As in equation~(\ref{eqn-pp11}), when the cloud optical depths
$\tau _a$ are much less than unity we can rewrite 
equation~(\ref{eqn-pp18}) as 
\begin{equation}
	\dt ^2_c \; = \; \int_0^{t_o}
	dt_idt_j\sigma ^2n_e(i)n_e(j) \; 
	{\lb \lb\tau v\rb_i \lb\tau v\rb_j \rb
        \over\lb\tau\rb_i \lb\tau\rb_j} \;
	e^{-\int_{k>i} \sigma n_edt_k}
	e^{-\int_{l>j} \sigma n_edt_l}\, .
\label{eqn-pp19}
\end{equation}
This is the expression one would write down in
perturbation theory based on the smoothed velocity and density
fields defined in equations~(\ref{eqn-pp5}), (\ref{eqn-pp6}), 
and (\ref{eqn-pp11}).
If $t_f$ is large compared to the scale of nonlinear 
clustering, this smoothed velocity field is well approximated by 
linear perturbation theory. 

\subsection{The shot noise contribution to $\delta T/T$}
\label{shot-noise}

We consider here some simplified forms for the shot noise term in
equation~(\ref{eqn-pp15}). We assume the universe is
ionized back to high redshift where the scattering optical depth
to the present is large. If the distributions of $\tau_a$ and
$v_b$ are independent of time, the integral is
\begin{equation}
	\dt ^2_s \; = \; 
	{\lb v^2(1-e^{-\tau })^2\rb\over 1-\lb e^{-2\tau }\rb } \, .
\label{eqn-sn1}
\end{equation}
We see that if $\tau\gg 1$ the temperature perturbation is just 
the rms velocity of the last cloud along the line of sight. 
If the dispersion in $\tau$ is small, equation~(\ref{eqn-sn1}) is 
\begin{equation}
	\dt ^2_s \; = \; 
	\lb v^2\rb {1-e^{-\tau }\over 1+e^{-\tau }} =
	\lb v^2\rb {q\over 2 - q}\, , 
\label{eqn-sn2}
\end{equation}
where the probability a photon is scattered by a cloud is
\begin{equation}
	q \; = \; 1 - e^{-\tau }\, .
\label{eqn-sn3}
\end{equation}
Another derivation of equation~(\ref{eqn-sn2}) is in \cite{pe90}.

In the numerical examples in \S 5 the cloud optical
depths $\tau$ are small. Here, following 
equations~(\ref{eqn-pp11}) and
(\ref{eqn-pp19}), we can write equation~(\ref{eqn-pp15}) as 
\begin{equation}
	\dt ^2_s \; = \; 
	\int_0^{t_o} \, dt \, {\lb\tau ^2v^2\rb\over\lb\tau\rb } \,
	\sigma n_e \,\exp -2\int _t^{t_o}\sigma n_edt'\, .
\label{eqn-sn4}
\end{equation}
This is the average across the Poisson process for a given
realization of the smoothed fields that represent the large-scale
structure. In linear perturbation theory the average 
across the smoothed fields just replaces the density and velocity
dispersion with the global mean values as functions of world time. 
One could numerically evaluate the resulting time integrals, 
but if $\lb\tau ^2 v^2\rb/\lb\tau\rb$ 
is not a rapidly varying function of time the integral is well
approximated by the value of this expression at the epoch of last
scattering. In this approximation the shot noise term is
\begin{equation}
	\dt ^2_s 
	\; = \; {\lb\tau^2 v^2\rb_s\over 2\lb\tau\rb_s}
	\; = \; {\tau_s v_s{}^2\over 2}\, .
\label{eqn-sn5}
\end{equation}
The factors in the last expression are suitably weighted
values of the mean cloud optical depth and mean square peculiar
velocity evaluated at the epoch of last scattering of the CBR,
under the assumption that the baryons are concentrated in clouds
back to this epoch.

To describe the effect of scattering at lower redshifts 
we use the mean scattering optical depth,
\begin{equation}
\bar \tau (z) \; = \; \int_{0}^z \,\sigma {\bar n_e} (dt/dz)dz ,
\label{eqn-sn6}
\end{equation}
where $\bar n_e$ is the cosmic mean density of free electrons in
optically thin clouds. In the ``saturated'' case, the integral in
equation~(\ref{eqn-sn6}) increases indefinitely
with $z$, and the redshift of last scattering, $z_s$, is defined
by $\bar\tau (z_s) = 1$. Equation
(\ref{eqn-sn5}) assumes this saturated ionization case. Another 
possibility is that $\bar\tau (z)$ does not 
grow indefinitely with $z$ but rather reaches a maximum
value $\bar\tau < 1$ at $z = z_{\star}$, and then remains constant
back to the hydrogen recombination epoch. 
Here the approximation to the shot
noise term in equation~(\ref{eqn-sn4}) changes from
equation~(\ref{eqn-sn5}) to 
\begin{equation}
\dt _s^2 \; = \; {\tau_s v _o^2\over 2}
\left( v_s\over v_o\right) ^2 \,
\left(1 - e^{-2 \, \bar\tau} \right) \; .  
\label{eqn-unsaturated}
\end{equation}
The plasma streaming velocity $v_s$ at last scattering, at
redshift $\sim z_{\star}$, 
has been scaled to the present value $v_o$ by the factor $v_s/v_o$
computed in linear perturbation theory. It will be recalled that
the typical optical depth per cloud is $\tau _s$. 

To compare our results with experimental data and theoretical
predictions for primary anisotropies, it is convenient to expand
the CBR temperature two-point angular correlation function in Legendre
polynomials, 
\begin{equation}
C(\Theta) \; = \; \lb\,{\textstyle
{\delta T\over T}(1){\delta T\over T}(2)} \,\rb \; 
= \; \sum_{\ell} \, {{2\ell + 1} \over 4\pi} \, 
C_{\ell} \, P_{\ell}\,(\cos \Theta),
\label{eqn-C}
\end{equation}
where $\Theta$ is the angle between directions (1) and (2),
while
\begin{equation}
C_{\ell} \; = \; \lb \, |a_{\ell}^m|^2 \, \rb \; , 
\label{eqn-alm}
\end{equation}
and $a_{\ell}^m$ is a coefficient in the spherical harmonic expansion
of the CBR temperature distribution.
The shot noise contribution to the CBR
temperature autocorrelation function approximates a step
function, $C(\Theta )= (\delta T/T)_s^2$ at angular separation
$\Theta$ smaller than the typical angular size $\theta _{\rm cl}$ of a
cloud at redshift $z_s$, and $C(\Theta ) = 0$ at 
$\Theta >\theta_{\rm cl}$. The angle $\theta _{\rm cl}$
subtended by a cloud of
diameter $d _{\rm cl}$ at redshift $z_s$ is
\begin{equation} 
\theta _{\rm cl} \; = \; H_o d _{\rm cl}(1+z_s)/y \; ,
\label{eqn-thetacl}
\end{equation}
where $y$ is the usual dimensionless angular size distance
(Peebles 1993, eq. [13.29]),
\begin{equation}
y(z_s,\Omega,\Lambda,H_o) \; = \; H_o \, a_o \, r \; ,
\label{eqn-y}
\end{equation}
with $r$ equal to the comoving angular size distance to a cloud
at $z_s$. 
To avoid ringing 
in $\ell$-space, we replace the step function-like $C(\Theta)$ 
by a Gaussian,
\begin{equation}
C(\Theta) = (\delta T/T)_s^2 \, \exp(-\Theta^2/\theta_{\rm cl}^2)
\; .
\label{eqn-ctheta}
\end{equation} 
If the shot noise dominates, and $\ell \gg 1$,
the resulting power spectrum of CBR 
fluctuations is given by the usual Hankel transform, 
\begin{equation}
C_{\ell} \; = \; 
\left({\textstyle {\delta T\over T}}\right)_s^2 \,\int_0^{\infty}
\exp(-\Theta^2/\theta_{\rm cl}^2) \, J_o (\Theta\ell) \,
2\pi \, \Theta \, d\Theta
\; = \; \pi\,\theta_{\rm cl}^2 \, 
\left({\textstyle {\delta T\over T}} \right)_s^2 \,
\exp( - \ell^2\theta_{\rm cl}^2/4) \; ,
\label{eqn-cls}
\end{equation}
where $J_o$ is the zeroth order Bessel function. In these models
the power per octave, $\sim \ell(\ell + 1)C_{\ell}$, rises with a growing
multipole number like $\ell^2$, reaching its peak value $\approx
(\delta T/T)_s^2$ near $\ell \approx \theta_{\rm cl}^{-1}$, and
then drops. This maximum signal is diluted in experiments with
antenna beamwidth greater than $\theta_{\rm cl}$. The mean   
square CBR temperature anisotropy averaged through a Gaussian
beam response of dispersion $\theta/2$, corresponding to full 
width at half maximum of $1.2 \theta$, is
\begin{equation}
\dt _{\theta}^2 \; \; = \; \;
{\theta_{\rm cl}^2 \over{\theta^2 + \theta_{\rm cl}^2}} \;
\dt _s^2
\; .
\label{eqn-dilute}
\end{equation} 
The image of the microwave sky produced with a
radiotelescope beam $\theta \gg \theta_{\rm cl}$ 
will be resolution limited, with the rms $\delta T/T$ reduced by
the factor $\theta _{\rm cl}/\theta$. Equations
(\ref{eqn-unsaturated}) and
(\ref{eqn-dilute}) are the analog of the Ostriker-Vishniac
expression in equation 
(\ref{eqn-ov1}), as we discussed in connection with equation
(\ref{eqn-ov5}).

\subsection{The contribution from correlated motions}

Here we consider the term induced by the correlations
in the peculiar velocity field. We ignore the correlations of 
cloud positions in our analysis for the following reason.

If the power spectrum of density fluctuations at zero redshift
is given by $P(k)$, where
$k$ is the wavenumber, then the linearized continuity
equation says that the present power spectrum of the peculiar
velocity field is $P_v(k) = H_o^2a_o^2f^2(\Omega, \Lambda)P(k)/k^2$, 
where $f \approx \Omega^{0.6}$ (e.g. \cite{pe93}, \S 13).  
Because of the factor $k^{-2}$ the coherence length of the density 
field is always significantly shorter than that of the velocity.
Moreover, density correlations tend to decrease more rapidly with
increasing redshift than velocity correlations. Thus in an
Einstein-de Sitter universe, 
$P(k,z) = P(k)(1+z)^{-2}$ while $P_v(k,z) =
P_v(k)(1+z)^{-1}$, so at high redshift
\begin{equation}
(v_z/v_o)^2 \; \equiv \; P_v(k,z)/P_v(k) \; \gg \; 
P(k,z)/P(k) \, .
\label{eqn-vz}
\end{equation}
A similar inequality holds for a wider class of models,
with $\Omega < 1$ and/or $\Lambda \ne 0$ (\cite{pe93},
\S13). For all models under consideration here,
the left-hand side
of the above inequality is larger than the right-hand side
by a factor $\sim \, z_s \gg 1$. In other words, 
the spatial correlations between the clouds can be safely ignored.

As in \S3.3, to simplify our calculations further, 
we now assume the dispersion in the cloud optical depths  is
small and that the probability 
of Thomson scattering per cloud is given by equation (\ref{eqn-sn3}),
where $q$ and $\tau$ do not depend on redshift (while in more 
realistic models $\tau$ may be a function of the cloud and the
redshift). Instead of the cloud coordinates
along the past light cone, $(t_i,t_j)$, it is more convenient
to use the mean and relative position, given by  
\begin{equation}
t = t(i,j) = (t_i + t_j)/2 \, , \qquad u = u(i,j) = t_i - t_j \, ,
\label{eqno-coord}
\end{equation}
as in the usual derivation of the Limber equation (see e.g.,
\cite{pe93}, \S7). After this coordinate transformation,  
relations (\ref{eqn-pp15}) and (\ref{eqn-pp18}) become
\begin{equation}
\dt _s^2 \; \; = \; \;
q^2 \int_0^{t_o} \, {dt \over t_f(t)} \, e^{-(2q - q^2)
\int_t^{t_o} dt'/t_{f}(t')}\, \, \Pi(0,t) \, ,
\label{eqn-pi0}
\end{equation} 
and 
\begin{equation}
\dt _c^2 \; = \;
2q^2(1-q)\int_0^{t_o} {dt\over t_f(t)} \, e^{- (2q - q^2)\int_t^{t_o}
dt'/t_f(t')} \, \int_0^{\infty} {du\over t_f(t)} \, e^{- uq/t_f(t)}
\, \Pi(u,t) \, ,
\label{eqn-piu}
\end{equation}
Here $\Pi$ is the radial component
of the velocity correlation tensor,
\begin{equation}
\Pi(u,t) \; = \; (v_z/v_o)^2 \, \Pi(r)
\label{eqn-radial}
\end{equation}
$u$ is the proper separation, 
$r = u/a(t)$ is the comoving separation, and 
$a(t)$ is the expansion parameter. 

The function $\Pi(r)$ can be expressed in terms
of the power spectrum,
\begin{equation}
\Pi(r) \; = \; {(a_oH_of)^2 \over 2{\pi^2}} \int_0^{\infty}
P(k) \, \left[ j_0 (kr) - 2 {j_1(kr)\over kr} \right] \,dk \, ,
\label{eqno-pir}
\end{equation}
where the $j_{\ell}$ are spherical Bessel functions
(\cite{gr89}, \cite{pe93}). At $r = 0$,
the value of the expression
in square brackets is 1/3. Hence,
the source term of the shot noise contribution (i.e., 
the one-dimensional velocity dispersion) is
\begin{equation}
\Pi(0,t) \; = \; \left({v_z \over v_o}\right)^2 \,
{(a_oH_of)^2 \over 6{\pi^2}} \int_0^{\infty} \, P(k) \, dk \, .
\label{eqn-shot}
\end{equation} 
In our analysis of the relative importance of the contribution
from correlated motion, this expression should be
compared with the inner integral in equation
(\ref{eqn-piu}), which can be written as
\begin{equation}
\int_0^{\infty}\,{du \over t_f} \, e^{-uq/t_f} \Pi(u,t) =
\left({v_z \over v_o }\right)^2 \, {(a_oH_of)^2 \over 2\pi^2 q} \,
\int_0^{\infty}\, P(k)W(kt_f/aq)\,dk \, ,
\label{eqn-corr}
\end{equation}
where the window function $W$ is
\begin{equation}
W(\kappa) = \kappa^{-1} \int_0^{\infty}\, dx \, e^{-x/\kappa}
\left[ j_0(x) - 2j_1(x)/x \right] \, .
\label{eqn-window}
\end{equation}
The variable $\kappa(k,t) = kt_f/aq$ is the ratio
the photon mean free path, $t_f(t)/q$ to the proper length 
$a(t)/k$ corresponding to the wavenumber $k$. 
The integral (\ref{eqn-window}) can be expressed in terms of
elementary functions (see appendix A),
\begin{equation}
W(\kappa)\, = \, \kappa^{-2} - \kappa^{-3} {\rm arctan}(\kappa) 
\, \approx \, (3 + \kappa^2)^{-1} \, .
\label{eqn-closedform}
\end{equation}
The last expression is exact at $\kappa = 0$ and in the limit
$\kappa \rightarrow \infty$. For any intermediate values
of $\kappa$, the accuracy of the
approximate expression is better than 20\%.  
The window function $W[\kappa(k,t)]$
acts as a low pass filter, damping the contribution to $(\delta T/T)_c$
induced by correlated motions on characteristic scales 
that are shorter than the photon mean free path.
Whether or not $(\delta T/T)_c$ can be neglected in comparison to
$(\delta T/T)_s$ thus depends on the ratio of the velocity
coherence length to the photon mean free path at $z_s$. 
To be more specific, let us model the power 
spectrum as
\begin{equation}
P(k) \; = \; {Ak \over (1 + k^2r_v^2)^{\beta}} \, ,
\label{eqn-toyspectrum}
\end{equation}
where $A, r_v$ and $\beta$ are constant parameters.
The APM data suggest $\beta = 1.2$ and $r_v = 33$ h$^{-1}$Mpc
\cite{ba96}. We trade precision for simplicity and 
for the purpose of our order of magnitude estimates set 
\begin{equation}
\beta \; = \; 1 \, .
\label{eqn-beta}
\end{equation}
To make this spectrum applicable for our linear theory
expression for $\Pi(r)$, we need a high wavenumber cutoff to
account for the stabilization of clustering by virialization,
occurring on comoving scales smaller than 
$k_{nl}^{-1} \ll r_v$, where the density contrast exceeds unity.
We therefore set $P(k) = 0$ for $k > k_{nl}$. Now the integrals
in equations (\ref{eqn-shot}) and (\ref{eqn-corr}) become trivial,
and the ratio
of the correlated source term to the shot noise source term is
\begin{equation}
{\int_0^{k_{nl}}\,P \,W \, dk \over \int_0^{k_{nl}} \,
P \, dk} \; \; = \; \;
{|\ln\left[(k_{nl}^2 \lambda^2 + 3)/3(k_{nl}^2 r_v^2 + 1)\right]|
\over q \, |(\lambda/r_v)^2 - 3 | \, \ln(1 + k_{nl}^2 r_v^2)} \; ,
\label{eqn-ratio}
\end{equation}
where $\lambda \equiv t_f/qa$,
and $r_v/\lambda$ is the ratio of the velocity coherence length
in proper coordinates, $r_va(z)$, to the  mean free path of
the CBR photons, $t_f(z)/q$. The correlated term becomes 
negligibly small when the condition
\begin{equation}
{1 \over q} \, \left({qr_v \over t_f(z_s)z_s }\right)^2 \; \;
\ll \; \; 1
\label{eqn-condition}
\end{equation}
is satisfied. Note that to make the correlated term dominant,
it is not enough to require that $r_va(z_s) \gg t_f(z_s)/q$;
the clouds have to be optically thin as well $(q \ll 1)$. 
The coherence of the velocity field can amplify
the net CBR temperature fluctuation by causing the effects of
individual clouds to add with the same sign and not as a random
walk. This, however, can happen only if
(1) the velocity coherence length is larger than the photon
mean free path and (2) the rescattering cutoff in the sum appears
sufficiently far away from the observer so that amplification by
coherent motions 
is not destroyed by rescattering. The second condition 
means good visibility, or $q \ll 1$. Indeed, when 
$r_v/\lambda \gg 1$, the ratio of the correlated term to the shot
noise term is 
\begin{equation}
{\int_0^{k_{nl}} \, P\,W \, dk \over \int_0^{k_{nl}} \,
P \, dk} \; \; \approx \; \; {\ln(\sqrt{3}r_v/\lambda)\over 3q\ln(k_{nl}r_v)} \; .
\end{equation}
Hence, unless the optical depth per cloud is small, the correlated term
will be subdominant, even in the large-$r_v$ limit.

\section{A simplified model}

We present here a simplified model that helps clarify the
physics behind the contributions to $\delta T/T$ from shot noise
and correlated motions of the gas clouds. The model assumes the
universe is static, each 
cloud has the same scattering probability $q = 1 - e^{-\tau}$
(eq.~[\ref{eqn-sn3}]), and the mean free path between clouds is
large enough that we can neglect the correlation in cloud
positions. The model takes account of the broader correlation of
peculiar velocities. We characterize the strength and range of
the velocity correlations by an `effective' correlation function,
\begin{equation}
\Pi(r) = \sigma_v^2 \, e^{-r/r_v} \; ,
\label{eqn-rv}
\end{equation}
where $r_v$ is the velocity coherence length and $\sigma_v^2 = \Pi(0)$
is the one-dimensional velocity dispersion. The assumption that
the correlation of cloud positions may be neglected means 
the probability distribution of relative separations
$r$ between pairs of clouds,
intersected along a line of sight, and numbered $i$ and $i+j$, 
is the Poisson expression
\begin{equation}
dp(r,j) \; = \; {(r/r_f)^{j-1} \over (j-1)!} \;
e^{-r/r_f}\; {dr\over r_f} \; ,
\label{eqn-poisson}
\end{equation}
where
\begin{equation}
r_f \; = \; \int_0^{\infty} \, r \, dp(r,1) 
\; = \; 4/ n\pi d_{cl}^2 \; , 
\label{eqn-rf}
\end{equation}
is the mean separation between clouds along the line of sight,
$n$ is the spatial density of clouds and $d_{cl}$ is the cloud
diameter and the cloud numbers increase with distance from the observer. 
The perturbation to the CBR along a given line of sight is (Peebles 1990)
\begin{equation}
\dtn \; = \;
q v_1 \, + \, q(1-q) v_2 \, + \, q(1-q)^2 v_3  \, + \, \ldots
\label{eqn-series}
\end{equation}
The line of sight component of the $j^{\rm th}$ velocity is
$v_j$, counting back from the present. 
The prefactor $q(1-q)^{j-1}$ is the probability $q$ that
a photon is scattered into our line of sight by the cloud $j$,
multiplied by the probability that it is not scattered out
of our line of sight by the remaining $j-1$ clouds in the foreground.  
If the universe is ionized back to large redshift
equation~(\ref{eqn-series}) in effect is a convergent infinite
series, meaning the influence of
consecutive terms decreases with increasing 
cloud number $j$. The
attenuation by rescattering out of the line of sight 
means the CBR perturbation is determined by the last 
$\sim N$ cloud intersections, where 
\begin{equation}
N \; \approx \; q^{-1} \; .
\label{eqn-N}
\end{equation}
The attenuation by rescattering is most severe 
when the clouds are opaque $(q = 1)$, and 
all $j > 1$ terms vanish. The nearest
cloud obscures the more distant ones, the anisotropy is
determined only by the velocity of the nearest cloud,
$\delta T/T = v_1$, and 
\begin{equation}
\lb \, \dt^2 \, \rb \; = \; \sigma_v^2 \; . 
\label{eqn-opaque}
\end{equation}
Let us now return to the more interesting case when $q < 1$.
Squaring the expression (\ref{eqn-series}), and averaging, we get
\begin{equation}
\lb \, \dt^2 \, \rb \; = \;
q^2 \, \sum_{i=1}^{\infty} \sum_{j=1}^{\infty}
(1-q)^{i+j-2} \lb \, \Pi \, \rb_{ij} \; ,
\label{eqn-altdT}
\end{equation}
where $\lb\,\Pi \, \rb_{ii}=\sigma _v^2$ and for $j>i$
the mean of the velocity autocorrelation function  
(eq.~[\ref{eqn-rv}]) for the Poisson distribution in
the distance between cloud intersections $i$ and $j$
(eq.~[\ref{eqn-poisson}]) is 
\begin{equation}
	\langle\Pi \rangle _{ij} = \sigma _v^2\int _0^\infty
	{du\over r_f}\left( u\over r_f\right) ^{j - i - 1}
	{1\over (j - i - 1)!} \; e^{-u(1/r_f + 1/r_v)}
	= {\sigma _v^2\over (1 + r_f/r_v)^{j - i}}. 
\end{equation}
With this expression the sums in equation~(\ref{eqn-altdT}) are
readily evaluated to get
\begin{equation}
\lb \,\dt^2 \, \rb \; = \; {\sigma _v^2q\over 2 - q}\,\left[\,
1 \, + \, {2(1-q)(r_v/r_f)\over 1+q(r_v/r_f)}\,\right]  \; .
\label{eqn-model}
\end{equation}
The first term in the square brackets comes from the sum of the
squared terms in equation~(\ref{eqn-altdT}) and the second from
the sum of the cross terms.  

If the velocity field is uncorrelated, $r_v = 0$, the
contribution from the cross terms vanishes and
equation~(\ref{eqn-model}) reduces to equation (\ref{eqn-sn2}).
The cross terms also vanish if the nearest cloud
along the line of sight is opaque, $q=1$, and the
anisotropy reduces to the ``single cloud'' limit
(\ref{eqn-opaque}), independent of the velocity correlation
length. The ``single cloud'' limit also applies when the velocity
correlation length is large, $qr_v\gg r_f$, because the photon
mean free path samples a single coherently moving set of clouds. 
And more generally, a positive
velocity correlation reduces the effective number of
statistically independent steps in the random 
walk and increases the rms temperature perturbation. 
 
We note finally that one can take account of the Hubble
expansion by replacing equation~(\ref{eqn-poisson})
with an inhomogeneous Poisson process,
\begin{equation}
dp(t) \; \; = \; \;{dt\over t_f(t)} \, 
{\left(\int_t^{t_o}\,dt'/ t_f(t')\right)^{i-1}\over(i-1)!}
\; e^{- \int_t^{t_o}\,dt'/t_f(t')} \; ,
\label{eqn-restore}
\end{equation}
from which it is an interesting exercise to rederive 
equations~(\ref{eqn-pi0}) and~(\ref{eqn-piu}).

\section{Numerical examples}

Since we know very little about the history of structure
formation at redshifts $z\gap 5$ a good strategy is to
consider examples of what might have happened and how it would
have affected observables such as the CBR temperature
fluctuations. To keep our discussion of possibilities simple and
definite we choose one set of cosmological parameters,  
\begin{equation}
	\Omega = 0.2\, ,\qquad \Omega _B=0.05, \qquad 
	h = 0.7 = H_o/100\hbox{ km s}^{-1}\hbox{ Mpc}^{-1}\, .
\label{eqn-ne1}
\end{equation}
The low value of $\Omega$ is in line with the observational evidence
(\cite{fr94}; \cite{pe97b}; \cite{per98}; and references therein), 
and cosmogonies with the early structure formation we have
assumed in this analysis seem to be most promising in a low
density cosmological model (\cite{pe97a}). To emphasize the
scattering effect we have adopted a baryon density parameter near
the high end of the range now under discussion (\cite{CST}).
Following \cite{nfw96} we assume that at any epoch matter is
concentrated in clouds with density contrast
\begin{equation}
	\delta\sim 200.\label{eqn-delta}
\end{equation}
Then a cloud mass $M$ fixes a characteristic cloud diameter, 
\begin{equation}
d_{\rm cl}= [6M/(\pi\bar\rho _s\delta )]^{1/3},
\end{equation}
where $\bar\rho _s$ is the mean mass density at the scattering epoch
$z_s$. A measure of the physical length scale of nonlinear mass
fluctuations is 
\begin{equation}
	d_{\rm nl} = {15\hbox{ Mpc}\over h(1+z_s)}
	\left(D_s\over D_o\right)^{2/(3+m)}\, .
\label{eqn-ne5}
\end{equation}
A sphere with this diameter contains roughly the cloud mass $M$. 
We have normalized to the rms fluctuation of galaxy counts,
$\delta N/N\sim 1$ in a sphere of diameter $15h^{-1}$~Mpc. 
The assumption that galaxies trace mass agrees with our low value
for the density parameter $\Omega$. 
Equation~(\ref{eqn-ne5}) is scaled by the power law power spectrum 
\begin{equation}
  P\propto k^m, \label{eqn-em}
\end{equation}
to the epoch $z_s$ using the linear perturbation theory growth
factor for the density contrast, $\delta\rho /\rho\propto D(t)$.
Finally, we take the present rms peculiar velocity to be
\begin{equation}
v_o=600\hbox{ km s}^{-1},
\end{equation}
and we scale this velocity with time by the linear perturbation
theory relation $v\propto a\, dD/dt$. 

We consider first the effect of scattering by clouds at epoch
$1+z_s=10$. In a cosmologically flat model the density and
velocity growth factors are 
\begin{equation}
1+z_s=10,\qquad D_s/D_o=0.14,\qquad v_s/v_o=0.49.
\end{equation}
For a mass characteristic of the central part of a large
galaxy the power spectrum index (eqs.~[\ref{eqn-ne5}],
[\ref{eqn-em}]), the model parameters are 
\begin{eqnarray}
M_{\rm cl} = 1\times 10^{11}M_\odot,\qquad
m = -1.4,\qquad r_f = 4\hbox{ Mpc}, \nonumber \\ 
d_{\rm cl} = 33\hbox{ kpc}, \qquad y = 2.5, 
\qquad \theta_{\rm cl} = 6.4\hbox{ arc sec} \; .
\label{eqn-numbers}
\end{eqnarray}
Here $r_f$ is the mean free path for intersection of clouds,
$d_{\rm cl}$ and $\theta_{\rm cl}$ are the cloud diameter and
the angle it subtends, finally, $y$ is the usual angle
size distance parameter. (For comparison, note that for $\Lambda = 0$ and
$z_s \gg \Omega^{-1}$, the angular size distance is
$y = 2/\Omega$.) The shot noise contribution to the 
CBR anisotropy (eqs.~[\ref{eqn-unsaturated}] and
~[\ref{eqn-dilute}]) is
\begin{equation}
\delta T/T(\theta =1\hbox{ arc min}) = 2\times 10^{-6}.
\label{eqn-more}
\end{equation}
We are assuming all baryons are in optically thin plasma, 
so the mean optical depth for scattering back to this redshift is
$\bar\tau =0.1$; the probability of scatternig
per cloud is $\tau_s = 0.004$. The contribution to the Compton-Thompson 
parameter in the CBR spectrum by the motions of the galaxies is 
$y_{\rm c}\sim\bar\tau v_s^2=1\times 10^{-7}$, well within the COBE
bound (\cite{COBE}; note that $v_s$ here is expressed in units
of the velocity of light; in our units $c = 1$). 

With the parameters in the above example the mean baryon density
within a cloud is $n\sim 0.06$ protons cm$^{-3}$, and the net
mass density is equivalent to about $0.2$ protons cm$^{-3}$. These
numbers  may not be unreasonable for a protogalaxy. The power law
index in equation~(\ref{eqn-numbers}) is comparable to that of the
most recent version of the isocurvature CDM model, 
$m\sim -1.8$ (\cite{pe97a}).
If the characteristic cloud mass is reduced to 
$M_{\rm cl} =1\times 10^{10}M_\odot$ 
with all other parameters the same we get $m = -1.8$ and 
$\delta T/T = 6\times 10^{-7}$ in a window of one arc minute
diameter. In an open universe (with zero cosmological
constant) with $M_{\rm cl} =1\times 10^{11}M_\odot$ and all other
parameters the same $\delta T/T$ is 1.5 times the value in
equation~(\ref{eqn-more}). 

The relative size of the contribution to $\delta T/T$ by
correlated motions is determined by equation~(\ref{eqn-ratio}). 
With the numbers in equation~(\ref{eqn-numbers}) the 
scattering probability in a cloud is $q=0.004$, so if the
parameter $q(r_v/z_sr_f)^2$ were of order unity the
comoving velocity coherence length would have to be
$r_v\sim 700$~Mpc, much larger than that suggested by the APM
power spectrum (eq.~[\ref{eqn-toyspectrum}]; Baugh \&\
Gazta{\~n}aga 1996) or even the most radical interpretation of
bulk flow observations. That is, $\delta T/T$ in our
model is dominated by the shot noise.

At a larger redshift in a cosmologically flat model,
\begin{equation}
1+z_s=30,\qquad D_s/D_o=0.047,\qquad v_s/v_o=0.28,
\end{equation}
the probability of scattering is $\tau_s = 0.007$ per cloud and
the mean optical depth is $\bar\tau =0.6$ if all baryons are in 
optically thin plasma, and a cloud model is
\begin{eqnarray}
M_{\rm cl} = 1\times 10^{9}M_\odot,\qquad
m = -1.4,\qquad r_f = 300\hbox{ kpc}, \nonumber \\
d_{\rm cl} = 2.4\hbox{ kpc}, \qquad y = 3.1, \qquad
\theta_{\rm cl} = 1.1\hbox{ arc sec}, \qquad   
\end{eqnarray}
and 
\begin{equation}
\delta T/T(\theta =1\hbox{ arc min}) = 5\times 10^{-7}.
\end{equation}
At still larger redshift in a cosmologically flat model,
\begin{equation}
1+z_s=50,\qquad D_s/D_o=0.028,\qquad v_s/v_o=0.22,
\end{equation}
we get $\tau_s = 0.004, \, \bar\tau =1$, and a cloud model is
\begin{eqnarray}
M_{\rm cl} = 1\times 10^7M_\odot,\qquad
m = -1.7,\qquad r_f = 40\hbox{ kpc}, \nonumber \\
d_{\rm cl} = 0.3\hbox{ kpc}, \qquad y = 3.3, 
\qquad \theta_{\rm cl} = 0.2\hbox{ arc sec} \; , 
\end{eqnarray}
and the contribution to the CBR anisotropy is still smaller,
\begin{equation}
\delta T/T(\theta =1\hbox{ arc min}) = 7\times 10^{-8}.
\end{equation}
Here the contribution to the Compton-Thompson parameter is 
$y_{\rm c}\sim 1\times 10^{-7}$. 

The temperature fluctuations in these examples are well below the 
measured bounds (\cite{re89}; \cite{fo93};
\cite{su93}; \cite{ch97}; \cite{pa97}; \cite{an94}). 
In Figure 1 we plot power spectra for CBR anisotropies for
the above three models, calculated from equation
~(\ref{eqn-cls}). One sees that in the models we
consider measurements that resolve the clouds 
should detect secondary temperature fluctuations as large as primary
anisotropies at their maximum.

\section{Discussion}
\label{discussion}

The simplifying assumption for this analysis is that structure
formation is so well advanced at the epoch of last scattering
of the CBR by free electrons that the mean distance $t_f$ between
intersections of clouds 
along a line of sight is large compared to the scale of nonlinear
mass fluctuations. This allows us to model the clouds as an 
inhomogeneous random Poisson process determined by the mass 
density and peculiar velocity fields smoothed through a window of 
width $t_f$, and it leads to the shot noise contribution 
to the small-scale CBR anisotropy in equation~(\ref{eqn-unsaturated}). 
This approach is motivated by the isocurvature CDM model for
structure formation (\cite{pe97a}), in which structure formation
could commence at decoupling at redshift $z\sim 1000$. 
A second important motivation has been to complement the 
usual perturbative analysis of the effect of the nonlinear growth
of small-scale structure. The similarity of results from the
perturbative (\cite{psco95}) and nonperturbative approaches 
leads us to believe we have reliable methods for estimating the
effect of early nonlinear structure formation on the
CBR anisotropy. 

The observations of young galaxies and the intergalactic medium
at $z\sim 3$ indicate a situation intermediate between the
perturbative and nonperturbative cases. The
damped Lyman-$\alpha$ systems contain a significant
baryon fraction, and the mean distance between intersections 
of these clouds is large (at $z=3$ it is longer than the Hubble
length). There also is a significant baryon fraction in the
Lyman-$\alpha$ forest, and these clouds have a relatively
short mean free distance. It is not unreasonable to speculate
that the situation at much larger redshifts similarly calls for a 
combination of the two approaches to the analysis of the 
angular distribution of the CBR. 

The CBR anisotropy produced by the Sunyaev-Zel'dovich (1970)
effect of the 
hot electrons in clusters of galaxies, which certainly is dominated by
the shot noise term, offers an important constraint on the epoch of
collection of the intracluster plasma. The evidence from the analysis
of Persi et al. (1995) is that this constraint does not
yet rule out the early structure formation picture. And our
conclusion from the numerical examples in \S 5 is that 
within presently known observational constraints structure
formation could have commenced when the universe was optically
thick to scattering of the CBR.  

\acknowledgments

We acknowledge useful discussions with Andrei Gruzinov,
Wayne Hu, Alexandre Refregier and Jan Wehr.
This work was supported in part at the California Institute of 
Technology by the Sherman Fairchild Distinguished Scholar 
Program, at Princeton  University by the US National Science 
Foundation, and by grants from the Polish Government (KBN grants
No. 2.P03D.008.13 and 2.P03D.004.13) and by the Poland-US Maria
Sk{\l}odowska-Curie Fund at Princeton University and University of
Pennsylvania at Philadelphia.

%\clearpage

\appendix
\section{The Thomson scattering window function}

The integral (\ref{eqn-window}) can be expressed in terms of two
hypergeometric functions (\cite{ry94}, hereafter GR,
eq. [6.621.1]),
\begin{equation}
\textstyle{
W(\kappa) = (1 - \kappa^2)^{-1/2} \left[ \,
F\left({1\over2}, {1\over2};{3\over2};{\kappa^2 \over 1+\kappa^2}\right)  
-{2\over3} F\left({1\over2}, {3\over2}; {5\over2}; {\kappa^2 \over
1 + \kappa^2}\right) \, \right] \, .
}
\label{eqn-hyper}
\end{equation}
One of the hypergeometric functions turns out to be an
elementary function in disguise (GR, eq. [9.121.26]):
\begin{equation}
\textstyle{
F\left({1\over2},{1\over2};{3\over2};z^2\right) \; = \;
\arcsin (z)/z \; ,
}
\label{eqn-arcsin}
\end{equation}
while the other can also be expressed in terms of elementary
functions by differentiating equation
(\ref{eqn-arcsin}), and using the standard recurrence
relation  (\cite{abxx}, eq. [15.2.7])
\begin{equation}
{\partial \over \partial z} \, \left[
(1 - z)^a F(a, b; c; z)\right] \; = \;
- {a(c-b) \over c} \, (1-z)^{a-1} F(a+1, b; c+1; z) 
\label{eqn-recurrence}
\end{equation}
for $(a, b, c) = (1/2, 1/2, 3/2)$. The final result of all this
is the expression (\ref{eqn-closedform}).

\vfill\eject

\clearpage

\psfig{file=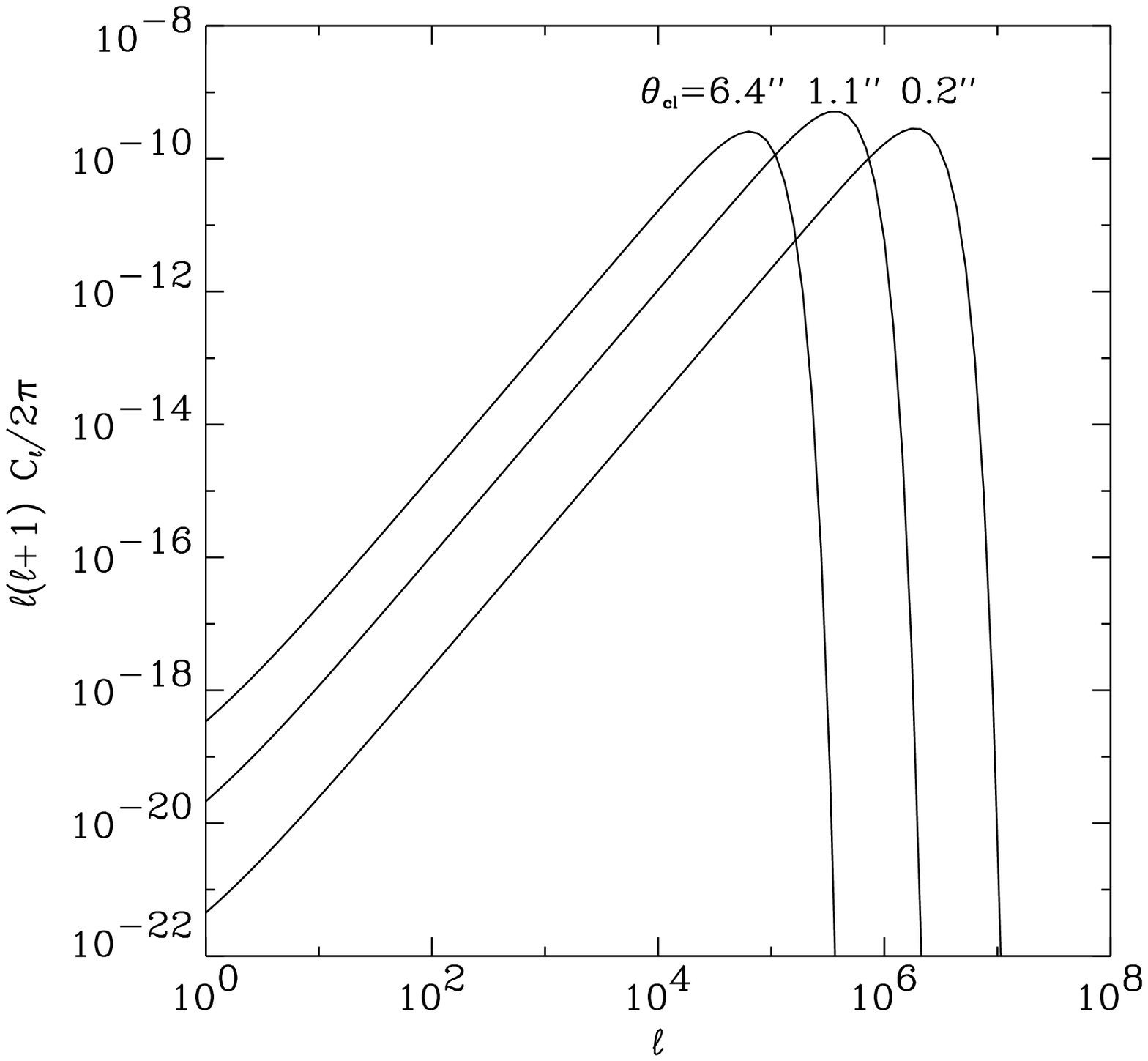,width= 1.0\textwidth}

%\noindent{\bf Figure caption}

\noindent{\bf Figure 1}

CBR angular power spectra, calculated from equations
(\ref{eqn-unsaturated}) and (\ref{eqn-cls}) for three sets of
model parameters, described in \S 5. The anisotropy is
induced by scattering in moving clouds of ionized hydrogen. 
To identify each of the three models, we label the power spectra
with the appropriate values of the cloud angular
sizes, $\theta_{\rm cl}$.

\end{document}